\documentclass[11pt]{article}
\usepackage[a4paper,left=30mm,top=40mm,bottom=40mm,right=30mm]{geometry} 
\usepackage{graphicx} 
\usepackage{cite}
\usepackage[utf8]{inputenc}
\usepackage[T1]{fontenc}
\usepackage{epstopdf}
\usepackage{ulem}
\usepackage{hyperref}
\usepackage{mathtools}
\usepackage{amsmath}
\usepackage{setspace} 
\usepackage{caption}
\usepackage{authblk}
\usepackage{newfloat}
\DeclareFloatingEnvironment[name={Supplementary Figure},fileext=lof]{suppfigure}
\usepackage[none]{hyphenat} 
\title{\textbf{Attosecond delays of high harmonic emissions from isotopes of molecular hydrogen measured by Gouy phase XUV interferometer}}

\author[1]{Mumta Hena Mustary}
\author[2,3]{Liang Xu}
\author[2]{Wanyang Wu}
\author[1]{Nida Haram}
\author[1]{Dane E. Laban}
\author[1]{Han Xu}
\author[2,4]{Feng He}
\author[1]{Igor V. Litvinyuk}
\author[1]{R. T. Sang}
\date{}
\affil[1] {Centre for Quantum Dynamics, Griffith University, Brisbane, Queensland 4111, Australia.}
\affil[2] {Key Laboratory for Laser Plasmas (Ministry of Education) and School of Physics and Astronomy,
Collaborative Innovation Center for IFSA (CICIFSA), Shanghai Jiao Tong University, Shanghai 200240, China.
}
\affil[3] {Shanghai Key Lab of Modern Optical System, University of Shanghai for
Science and Technology, 200093 Shanghai, People's
Republic of China.}
\affil[4] {CAS Center for Excellence in Ultra-intense Laser Science, Shanghai, 201800, China.}

\begin{document}
\maketitle

\textbf{High harmonic spectroscopy\cite{Smirnova2009,Mcfarland2009,Peng2019} can access structural and dynamical information on molecular systems encoded in amplitude and phase of high harmonic generation (HHG) signals\cite{Kanai2005,Li2008, Baker2006, Baker2008, Zair2013, Itatani2004}. However, measurement of the harmonic phase is a daunting task. Here we present a precise measurement of HHG phase difference between two isotopes of molecular hydrogen using the advanced extreme-ultraviolet (XUV) Gouy phase interferometer\cite{Mustary2018}. The measured phase difference is about 200 mrad, corresponding to $\sim$ 3 attoseconds (1 as = 10$^{-18}$ s) time delay which is nearly independent of harmonic order. The measurements agree very well with numerical calculations of a four-dimensional time-dependent Sch{\"o}dinger equation. Numerical simulations also reveal the effects of molecular orientation and intra-molecular two-centre interference on the measured phase difference. This technique opens a new avenue for measuring the phase of harmonic emission for different atoms and molecules. Together with isomeric or isotopic comparisons it also enables the observation of subtle effects of molecular structures and nuclear motion on electron dynamics in strong laser fields.}\\

\begin{figure*}[ht!]
\centering
\includegraphics[scale=0.3]{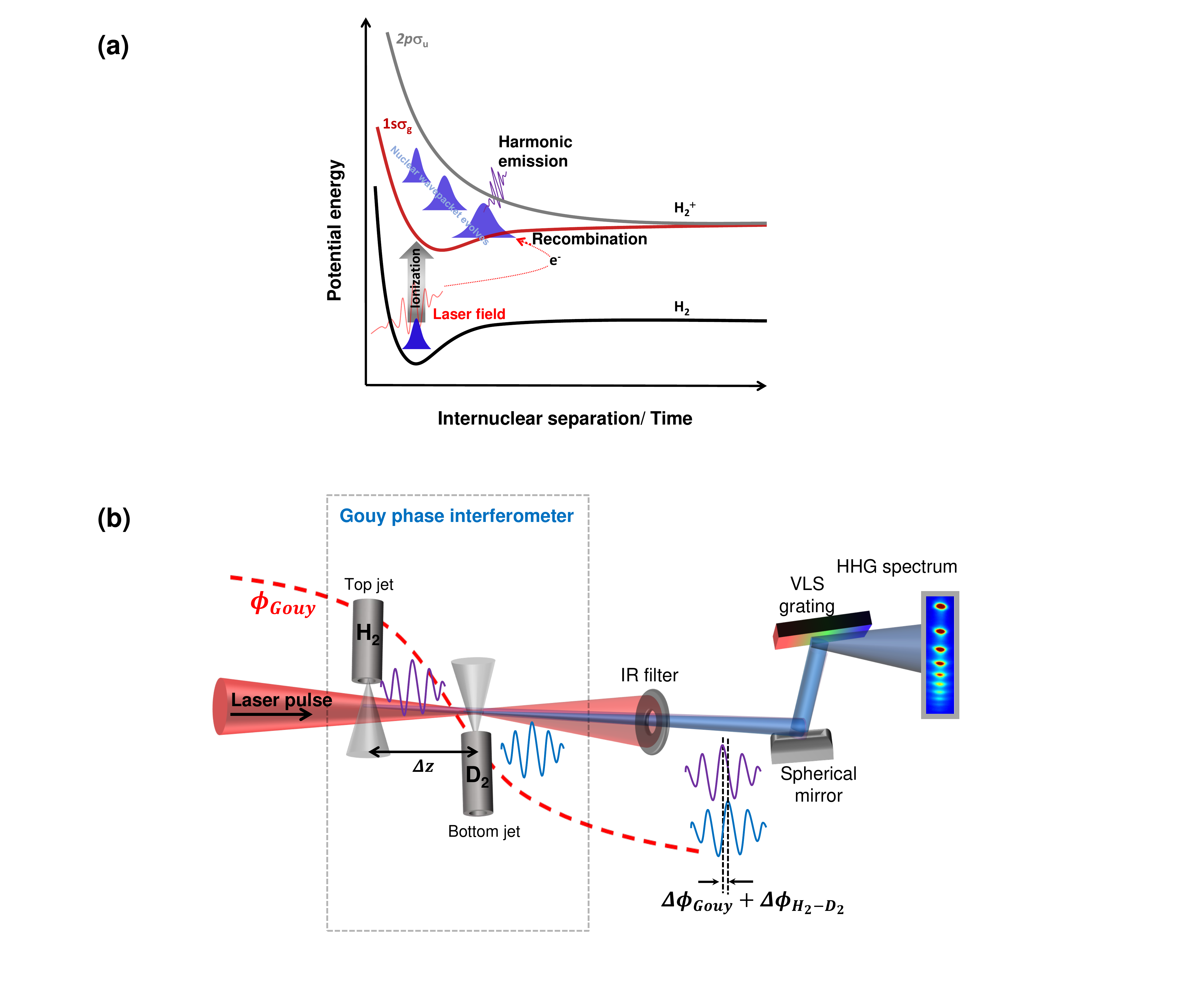}
\caption {\small{\textbf{(a) Three step model of HHG with the potential energy surfaces of H$_2$ molecule.} Tunneling ionization launches an electron wavepacket together with a nuclear vibrational wavepacket. The ionized electron is driven by the laser electric field and it returns to the parent ion at a certain recollision time. Simultaneously, the nuclear wavepacket evolves on the potential energy surface of the ground state of H$_2^+$. Upon electron recombination a photon is emitted with an amplitude and phase dependent on the correlated electron-nuclear dynamics.} \textbf{(b) Schematics of the Gouy phase interferometric technique.} Two gas jets are placed sequentially with a separation $\Delta z$ in a single laser focus. Two coherent high harmonic pulses are generated simultaneously from the top (H$_2$) and bottom (D$_2$) gas jets.  A phase difference between the two pulses includes contributions from the Gouy phase shift of the driving laser field at the two jets positions and the intrinsic phase difference between the two species. By measuring HHG yields for different gas configurations and jet separations these two contributions can be disentangled.}
\label{Fig 1.pdf}
\end{figure*}

 The HHG process is a sensitive probe of molecular dynamics and structures \cite{Biswas2020}. The hydrogen molecule, being the lightest, exhibits the fastest nuclear motion. Also, being the simplest neutral molecule, it allows accurate \textit{ab initio} quantum mechanical simulations without resorting to severe approximations. Those simulations can be used to understand and validate the experimental results. Moreover, availability of different isotopes adds isotopic comparison as a valuable benchmarking and validation tool. A number of studies with such comparisons have already been performed, including effects of nuclear dynamics on relative HHG yields and tunnel ionization rates in H$_2$ and D$_2$. The strong field ionization of H$_2$ launches a free electron with H$_2^+$ in the $1s\sigma_g$ state. The electron wavepacket driven by the laser field accelerates away from the parent ion. At the same time, the nuclear wavepacket evolves on the potential energy surface of H$_2^+$ (see Fig. \ref{Fig 1.pdf}(a)). After a certain time delay, the electron wavepacket returns to the parent ion upon the reversal of the driving laser field. Its consequent recombination induces an oscillating dipole, which imprints its amplitude and phase on the emitted HHG photons.\\

The investigation of the nuclear dynamics of hydrogen isotopes was envisioned theoretically {\cite{Lein2005, Bian2014, He2018}} and implemented experimentally\cite{Baker2006, Baker2008, Lan2017} by comparing their high-harmonic yields. However, only two studies explored the isotope effects on the high-harmonic phase\cite{Haessler2009,Kanai2008}. The first one estimated the relative phase by measuring harmonic yields in a mixed gas cell with H$_2$ and D$_2$\cite{Kanai2008}. The second study relied on the determination of group delay by using the reconstruction of attosecond beating by interference of two-photon transitions (RABITT) \cite{Haessler2009}. However, ensuring that both isotopes are at the same number density in the HHG interaction region is very challenging.   Additionally, a sign ambiguity\cite{Kanai2008} and  large experimental uncertainties of these measurements made it difficult to determine the phase difference and absolute delays accurately\cite{Haessler2009,Kanai2008}. Neither of those studies actually reproduced the measured phase difference theoretically. \\

To retrieve the phase information, the high-harmonic interferometers either split the driving laser beam into two paths and focus at different locations within a single gas jet that produce two phase-locked HHG photons \cite{Bellini1998, Lynga1999}, or use a single driving laser beam in a gas mixture \cite{Mcfarland2009, Kanai2008}. The first method is limited to a single gas species, while the second has low resolution and issues with determining the sign of the relative harmonic phase. Recently, the phase difference between two atomic species was measured by an all-optical attosecond interferometric technique\cite{Azoury2019}, where the delay between two extreme-ultraviolet (XUV) pulses is controlled by a two-segment mirror which can provide the temporal resolution of $\sim$ 6 attoseconds. Such an interferometer must maintain sub-cycle stability of its path difference (displacement between the two mirror segments) for XUV frequencies for the duration of the experiment while this path difference is being scanned. This is an exceedingly difficult task which severely limits practical utility of the device.\\

Building an interferometer in the XUV region is quite challenging for two reasons: firstly, it is challenging to control the delay of the XUV pulses precisely between the two arms with sub-cycle precision; secondly, the highly reflective XUV optics is yet to be developed. Our passively stabilized Gouy phase interferometer \cite{Dane2012,Mustary2018}, on the other hand, is an all-optical direct XUV interferometric technique. It does not require calibration of gas pressures to ensure the same number densities. Additionally, it does not require any XUV optics. The technique provides an elegant way to generate two coherent high harmonic pulses without splitting the driving laser and XUV beams. The two mutually coherent XUV pulses are generated by exploiting the inherent properties (Gouy phase) of a single Gaussian focused laser beam. Its unprecedented resolution of $\sim$ 300 $\mu $rad ($\sim$ 100 zeptoseconds) is a result of any instability in the distance between the two arms (the gas jets in this case) of the interferometer is determined relative to the Rayleigh length, $z_R$ of the fundamental laser beam as opposed to the XUV wavelength in a conventional optical interferometer \cite{Mustary2018}.\\

Here, we apply the technique to investigate the effect of nuclear dynamics on the electron motion in molecular hydrogen by precise measurement of high harmonic phase difference (and corresponding HHG phase delays) produced in H$_2$ and D$_2$. Since the ionization potentials of H$_2$  {($I_p$ = 15.43 eV)} and D$_2$  {($I_p$ = 15.46 eV)} 
are almost identical \cite{Baker2006}, the difference in the phase accumulated by electron in the continuum is considered to be negligible. However, due to the nuclear mass difference, the evolution of nuclear wave packet while electron propagates in the continuum and then recombines to the ground state may differ substantially. The harmonic intensity from the heavy isotope was shown to be higher compared to the lighter molecule as harmonic emission is sensitive to the nuclear motion \cite{Baker2006} though for D$_2$ the ionization probability is less than for H$_2$ \cite{Wang2016}. The aim of this work is to measure a small phase difference of HHG signals and to gain an insight into the correlated electron–nuclear dynamics for the two isotopes of molecular hydrogen.\\

The high harmonics are generated by a linearly polarized, 800 nm, 9 fs pulses with a peak intensity of $\sim 5\times 10^{14}$ W/cm$^2$. 
Our advanced Gouy phase interferometer consists of two spatially separated gas jets along the propagation direction in a single laser focus. One of the gas jets (bottom) is fixed in position at the center of laser focus while the second jet (top) can move along the laser propagation direction before the focus of the laser beam, as shown in Fig. \ref{Fig 1.pdf}(b) (see Methods for details). 
The Gouy phase modulates the carrier envelope phase (CEP) of a focused laser pulse which results in a phase shift of XUV pulses generated from the two gas jets given by

\begin{equation}
\begin{split}
\Delta \phi& = q\Delta \phi_{Gouy}\\
& =- q \tan^{-1} \Big(\dfrac{
\Delta z}{z_R}\Big),
\end{split}
\end{equation}

\noindent where $q$ is the harmonic order and $\Delta\phi_{Gouy}$ is the Gouy phase difference at two gas jets positions (separated by $\Delta z$). The diameter of the gas jets is 200 $\mu$m and gas pressure in both jets is kept at 100 Torr even though optimal phase matching happens at a slightly higher pressure. The short interaction region and low gas density help to reduce the macroscopic phase matching effects and to minimize the reabsorption of XUV photons by the second gas jet. The gases are switched in the jets by electronically actuated microvalves. That allows faster measurement and thus minimizes the errors due to laser fluctuation over time and also compensates for a small difference of intensities (and associated phases) at the two jets positions.\\

\begin{figure*}[ht!]
    \centering
    \includegraphics[scale=0.30]{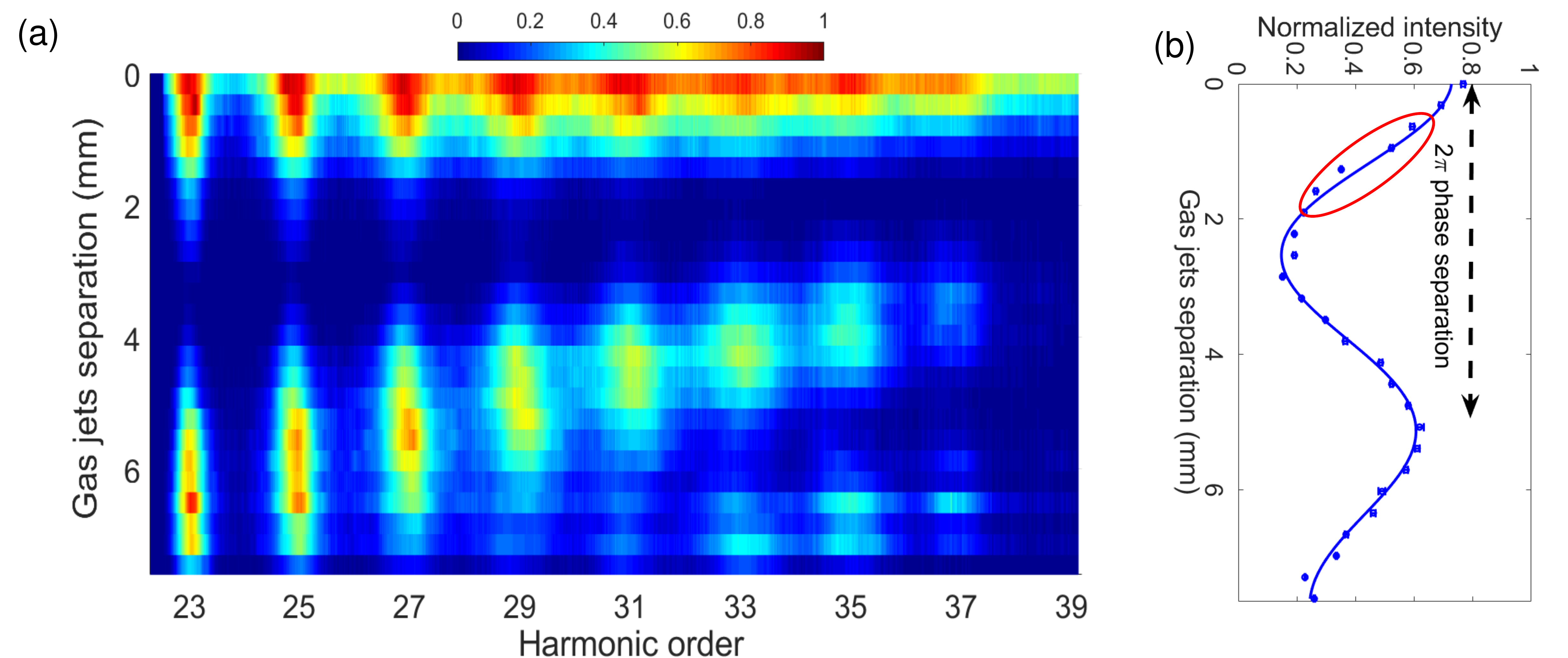}
    \caption{\small{\textbf{(a)} Normalized HHG yield from H$_2$ gas in both jets as a function of the jet separation. The two XUV pulses generated by a single laser pulse interfere constructively or destructively depending on the distance between the gas jets. This phase shift originates from the Gouy phase and it follows the relationship, $\Delta \phi=q \Delta \phi_{Gouy}= -q \tan^{-1} ({\Delta z}/{z_R})$, where $\Delta z$ is the separation between the jets and $z_R$ is the Rayleigh length of the driving laser beam. \textbf{(b)} The normalized yield of H27 as a function of jet separation. The area highlighted by the red ellipse represents the region of the highest resolution of the interferometer, i.e. where the largest intensity variation occurs for a given phase shift.}}
    \label{Fig 2.pdf}
\end{figure*}

The interference fringes of the high harmonics from H23 to H37 generated with H$_2$ in both jets obtained by varying the separation between the jets are shown in Fig. \ref{Fig 2.pdf}(a). 
To optimize the resolution for all the observable harmonic orders, the isotopic phase difference measurements were performed at jet separations of 0.63 mm and 1.27 mm where the fluctuation of laser intensities is only 0.2\% and 0.8\% respectively.\\

To extract the relative phase difference between the harmonics generated from the two gases, for each of the two jet separations measurements are performed for two configurations. In the first configuration, D$_2$ and H$_2$ are delivered from the top and bottom jets respectively. The gases are swapped in the second configuration.\\

\begin{figure*}[ht!]
   \centering
    \includegraphics[scale=0.28]{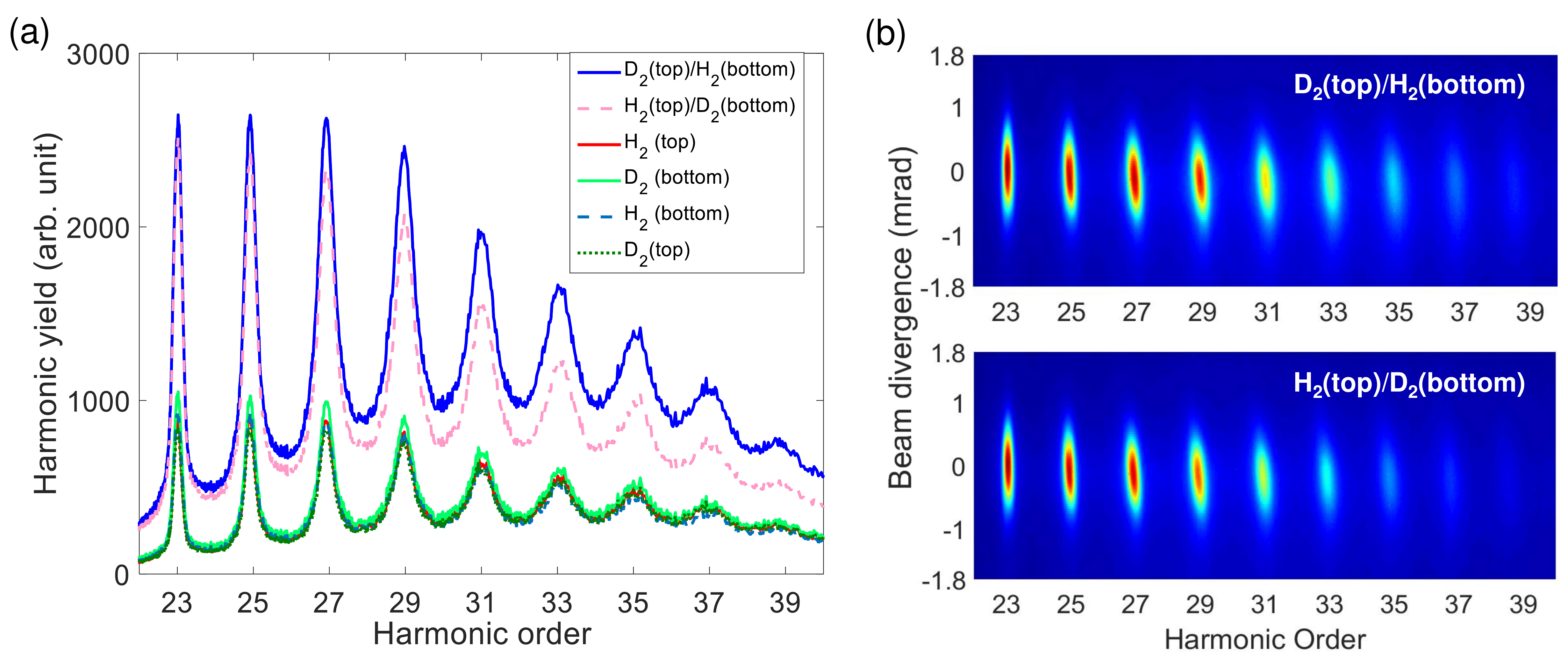}
    \caption{\small{\textbf{(a)} HHG spectra for the six different gas configurations used to determine the intrinsic relative phase difference between H$_2$ and D$_2$ at 0.63 mm separation between the gas jets. Each spectrum is an average of 200 images and each image is an integrated image of 100 laser shots.  \textbf{(b)} Images of the top two spectra in (a) illustrating the change due to switching the order of the gases.}}
    \label{fig:Phase_shift}
\end{figure*}

\begin{figure*}[ht!]
    \centering
    \includegraphics[scale=0.30]{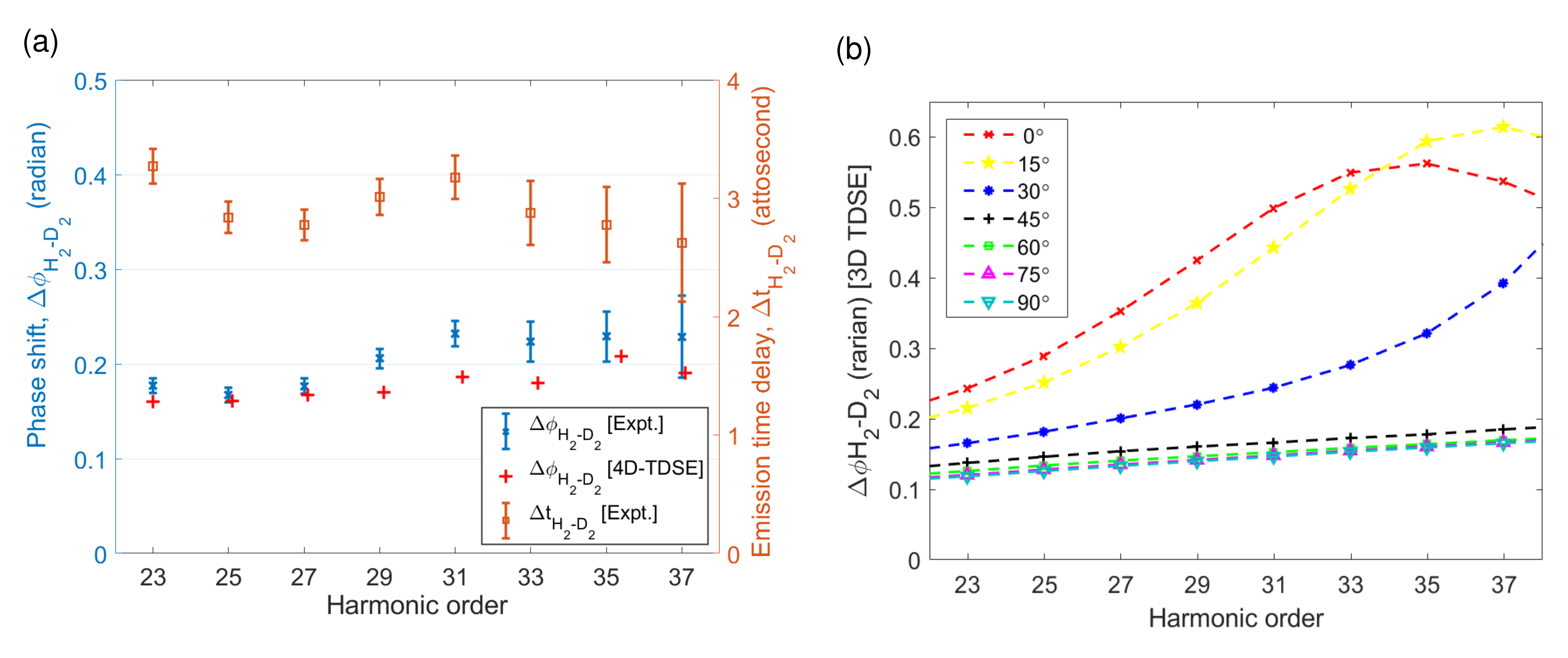}
    \caption{\small{\textbf{(a)} The phase differences (left vertical axis, blue) and phase delays (right vertical axes, red) between H$_2$ and D$_2$ for harmonics H23 to H37. The crosses represent theoretical results calculated by numerical 4D-TDSE model (see text for details). \textbf{(b)} The phase differences between H$_2$ and D$_2$ calculated by rotation-free 3D-TDSE model for molecules aligned at different angles $\theta_r$ with respect to the laser polarization. The plots for $\theta_r$ = 60, 75 and 90 degrees are on top of each other.}}
    \label{Fig4.pdf}
\end{figure*}

The HHG spectra are recorded for two different jet separations and for two gas configurations to account for systematic uncertainties. For each configuration, three HHG spectra are recorded: for both gas jets activated and for individual jet activated. The measured spectra are presented in  Fig. \ref{fig:Phase_shift}. The relative phase difference between H$_2$ and D$_2$ is calculated as

\begin{equation}
\Delta \phi_{H_2-D_2}={\rm sin}^{-1}\Bigg[\dfrac {I_{H_2}+I_{D_2}+2\sqrt{I_{H_2}I_{D_2}}}{4\sqrt{I_{H_2}I_{D_2}}}\Big(\dfrac{\Delta I_N}{sin(q\Delta \phi_{Gouy})}\Big)\Bigg],
\label{Eq:H2D2 phase shift}
\end{equation}


\noindent where, $I_{H_2}$ and $I_{D_2}$ are the harmonic intensities from individual gas jets of H$_2$ and D$_2$ respectively, and $\Delta I_N$ is the difference in the normalized intensity obtained from the two gas configurations. The results are presented in Fig. \ref{Fig4.pdf}(a). The left axis corresponds to the phase difference averaged over both separations and both gas configurations. It should be noted, that our method measures both the value and sign of the phase difference and the two different configurations of gases induce opposite signs. Their absolute values were averaged and the shown phases correspond to emission from H$_2$ being delayed relative to D$_2$ as our measurements indicate (see Methods for details).
The corresponding phase delays of harmonics from H$_2$ relative to D$_2$ are depicted on the right axis of Fig. \ref{Fig4.pdf}(a) and are determined by

\begin{equation}
\Delta t_{H_2-D_2}=\dfrac{\Delta \phi_{H_2-D_2}}{\omega_q},
\end{equation}
where, $\omega_q$ is the angular frequency of the $q^{th}$ harmonic order. The harmonics from H$_2$ are found to be $\sim$ 3 attoseconds delayed in phase with respect to D$_2$ for all the observable harmonic orders.\\

The experimental measurements are supported by numerical solutions of the four-dimensional time-dependent Schr{\"o}dinger equation (4D-TDSE) within the single active electron (SAE) approximation (see Methods for details). In this model, the electronic and nuclear motions are both confined to a plane containing the molecular vibration and rotation. The simulated results (red crosses) are presented in Fig. \ref{Fig4.pdf}(a) and they agree well with the experimental data (blue stars). Additional simulations showed that $\Delta \phi_{H_2-D_2}$ is relatively insensitive to the laser intensity, so the laser focal-volume averaging was not included. \\

 According to the well-known three-step model, the total HHG phase includes contributions from the ionization, propagation, and recombination processes. Thus, for molecules aligned along different angles $\theta_r$ towards the laser polarization direction, the dynamics of ionization and recombination should be different, resulting in the $\theta_r$-dependent $\Delta \phi_{H_2-D_2}$. Numerically, by excluding molecular rotations and restricting the nuclear motion to a single vibrational degree of freedom we reduced our 4D-TDSE model to a three-dimesional one. We used that 3D-TDSE model to obtain $\theta_r$-dependent  $\Delta \phi_{H_2-D_2}$, presented in Fig. \ref{Fig4.pdf}(b). 
 Only the molecules aligned at small angles (0-30 degrees) from the laser polarisation direction exhibit significant dependence of $\Delta \phi_{H_2-D_2}$ on harmonic order (energy of recolliding electrons) which could be attributed to two-center interference in the recombination step. Due to the different nuclear masses, the molecular bond elongation between the ionization and recombination steps is different for H$_2$ and D$_2$. In the recombination step, the complex two-center destructive interference occurs at a specific internuclear distance $R$ when the recombining electron momentum $k$ meets the condition $kR$cos($\theta_r)=\pi$, and the HHG phase undergoes a sudden change at the corresponding frequency $\omega=k^2/2+I_p$. It
 is the difference in two-center interference between H$_2$ and D$_2$ that is responsible for the angle and energy dependence of the curves in Fig. \ref{Fig4.pdf}(b). A more detailed analysis based on the Lewenstein model and  the two-center interference can be found in the supplementary information.\\

In summary, we have demonstrated a novel all-optical Gouy phase interferometric technique for measuring HHG phase difference between two atomic or molecular species. We used this technique to measure the HHG phase difference between H$_2$ and D$_2$ to be about 0.2 radian, corresponding to 3 attosecond phase delay for all the harmonics. We also simulated this phase difference by numerically solving the time-dependent Schr{\"o}dinger equation. The theoretical results obtained at the highest level of simulations agree quite well with the experiment. The simulations reveal that $\Delta \phi_{H_2-D_2}$ depends on the molecular alignment, bond stretching, and the two-center interference. The phase difference of harmonics from molecular isotopes can be used as a sensitive probe of ultrafast correlated electron-nuclear dynamics in molecules.\\

\textbf{Authors contribution:} D. L, I. L. and R. S. conceived, planned and lead the project. M. M and D. L. set up the interferometer and the experiments. M. M. carried out the measurement, analyzed and interpreted experimental data, improved the analysis and interpretation in discussion with D. L., I. L and R. S.. M. M wrote up experimental parts of this manuscript.  L. X., F. H and W. U conducted numerical simulations. L. X and F. H wrote up the theoretical sections. N. H and H. X. contributed to important discussion on results and theoretical analysis. All authors contributed in discussions on the results and prepared the manuscript for submission.\\

\textbf{Acknowledgements:} The experimental research was supported by the Australian Research Council (LP140100813 and DP190101145) and Griffith University. M. M. was supported by a Griffith University International Postgraduate Research (GUIPRS) and a Griffith University Postgraduate Research (GUPRS) Scholarships. Theoretical work was supported by National Key R\&D Program of China (2018YFA0404802, No. 2018YFA0306303), Innovation Program of Shanghai Municipal Education Commission (2017-01-07-00-02-E00034), National
Natural Science Foundation of China (NSFC) (Grant No. 11574205,  No. 91850203)

\bibliography{References}

\begin{thebibliography}{10}

\bibitem{Smirnova2009}
O.~Smirnova, Y.~Mairesse, S.~Patchkovskii, N.~Dudovich, D.~Villeneuve,
  P.~Corkum, and M.~Y. Ivanov.
\newblock High harmonic interferometry of multi-electron dynamics in molecules.
\newblock {\em Nature}, 460(7258):972, 2009.

\bibitem{Mcfarland2009}
B.~K. McFarland, J.~P. Farrell, P.~H. Bucksbaum, and M.~G\"uhr.
\newblock High-order harmonic phase in molecular nitrogen.
\newblock {\em Phys. Rev. A}, 80:033412, 2009.

\bibitem{Peng2019}
P.~Peng, C.~Marceau, and D.~M. Villeneuve.
\newblock Attosecond imaging of molecules using high harmonic spectroscopy.
\newblock {\em Nat. Rev. Phys.}, 1:144, Feb 2019.

\bibitem{Kanai2005}
T.~Kanai, S.~Minemoto, and H.~Sakai.
\newblock Quantum interference during high-order harmonic generation from
  aligned molecules.
\newblock {\em Nature}, 435(7041):470, 2005.

\bibitem{Li2008}
Wen Li, X.~Zhou, R.~Lock, S.~Patchkovskii, A.~Stolow, H.~C. Kapteyn, and M.~M.
  Murnane.
\newblock {Time-Resolved Dynamics in N2O4 Probed Using High Harmonic
  Generation}.
\newblock {\em Science}, 322(5905):1207--1211, 2008.

\bibitem{Baker2006}
S.~Baker, J.~S. Robinson, C.~A. Haworth, H.~Teng, R.~A. Smith, C.~C. Chiril{\u
  a}, M.~Lein, J.~W.~G. Tisch, and J.~P. Marangos.
\newblock Probing proton dynamics in molecules on an attosecond time scale.
\newblock {\em Science}, 312(5772):424--427, 2006.

\bibitem{Baker2008}
S.~Baker, J.~S. Robinson, M.~Lein, C.~C. Chiril\ifmmode~\u{a}\else \u{a}\fi{},
  R.~Torres, H.~C. Bandulet, D.~Comtois, J.~C. Kieffer, D.~M. Villeneuve,
  J.~W.~G. Tisch, and J.~P. Marangos.
\newblock Dynamic two-center interference in high-order harmonic generation
  from molecules with attosecond nuclear motion.
\newblock {\em Phys. Rev. Lett.}, 101:053901, 2008.

\bibitem{Zair2013}
A.~Za{\"i}r, T.~Siegel, S.~Sukiasyan, F.~Risoud, L.~Brugnera, C.~Hutchison,
  Z.~Diveki, T.~Auguste, J.~W.~G. Tisch, Pascal. Sali{\'e}res, M.~Y. Ivanov,
  and J.~P. Marangos.
\newblock Molecular internal dynamics studied by quantum path interferences in
  high order harmonic generation.
\newblock {\em Chem. Phys.}, 414:184 -- 191, 2013.

\bibitem{Itatani2004}
J.~Itatani, J.~Levesque, D.~Zeidler, H.~Niikura, H.~P{\'e}pin, J.~C. Kieffer,
  Paul~B. Corkum, and D.~M. Villeneuve.
\newblock Tomographic imaging of molecular orbitals.
\newblock {\em Nature}, 432(7019):867, 2004.

\bibitem{Mustary2018}
M.~H. Mustary, D.~E. Laban, J.~B.~O. Wood, A.~J. Palmer, J.~Holdsworth, I.~V.
  Litvinyuk, and R.~T. Sang.
\newblock Advanced gouy phase high harmonics interferometer.
\newblock {\em Journal of Physics B: Atomic, Molecular and Optical Physics},
  51(9):094006, apr 2018.

\bibitem{Biswas2020}
Shubhadeep Biswas, Benjamin F{\"o}rg, Lisa Ortmann, Johannes Sch{\"o}tz,
  Wolfgang Schweinberger, Tom{\'a}{\v{s}} Zimmermann, Liangwen Pi, Denitsa
  Baykusheva, Hafiz~A Masood, Ioannis Liontos, et~al.
\newblock Probing molecular environment through photoemission delays.
\newblock {\em Nature Physics}, pages 1--6, 2020.

\bibitem{Lein2005}
M.~Lein.
\newblock Attosecond probing of vibrational dynamics with high-harmonic
  generation.
\newblock {\em Phys. Rev. Lett.}, 94:053004, 2005.

\bibitem{Bian2014}
Xue-Bin Bian and Andr\'e~D. Bandrauk.
\newblock Probing nuclear motion by frequency modulation of molecular
  high-order harmonic generation.
\newblock {\em Phys. Rev. Lett.}, 113:193901, Nov 2014.

\bibitem{He2018}
Lixin He, Qingbin Zhang, Pengfei Lan, Wei Cao, Xiaosong Zhu, Chunyang Zhai,
  Feng Wang, Wenjing Shi, Muzi Li, Xue-Bin Bian, et~al.
\newblock Monitoring ultrafast vibrational dynamics of isotopic molecules with
  frequency modulation of high-order harmonics.
\newblock {\em Nature communications}, 9(1):1--7, 2018.

\bibitem{Lan2017}
P.~Lan, M.~Ruhmann, L.~He, C.~Zhai, F.~Wang, X.~Zhu, Q.~Zhang, Y.~Zhou, M.~Li,
  M.~Lein, and P.~Lu.
\newblock Attosecond probing of nuclear dynamics with trajectory-resolved
  high-harmonic spectroscopy.
\newblock {\em Phys. Rev. Lett.}, 119:033201, 2017.

\bibitem{Haessler2009}
S.~Haessler, W.~Boutu, M.~Stankiewicz, L.~J. Frasinski, S.~Weber, J.~Caillat,
  R.~TaÃ¯eb, A.~Maquet, P.~Breger, P.~Monchicourt, B.~Carr{\'{e}}, and
  P.~Sali{\`{e}}res.
\newblock Attosecond chirp-encoded dynamics of light nuclei.
\newblock {\em Journal of Physics B: Atomic, Molecular and Optical Physics},
  42(13):134002, jun 2009.

\bibitem{Kanai2008}
T.~Kanai, E.~J. Takahashi, Y.~Nabekawa, and K.~Midorikawa.
\newblock Observing the attosecond dynamics of nuclear wavepackets in molecules
  by using high harmonic generation in mixed gases.
\newblock {\em New Journal of Physics}, 10(2):025036, feb 2008.

\bibitem{Bellini1998}
M.~Bellini, C.~Lyng\aa{}, A.~Tozzi, M.~B. Gaarde, T.~W. H\"ansch,
  A.~L'Huillier, and C.-G. Wahlstr\"om.
\newblock Temporal coherence of ultrashort high-order harmonic pulses.
\newblock {\em Phys. Rev. Lett.}, 81:297--300, 1998.

\bibitem{Lynga1999}
C.~Lyng\aa{}, M.~B. Gaarde, C.~Delfin, M.~Bellini, T.~W. H\"ansch,
  A.~L'~Huillier, and C.-G. Wahlstrom.
\newblock Temporal coherence of high-order harmonics.
\newblock {\em Physical Review A}, 60:4823--4830, 1999.

\bibitem{Azoury2019}
Doron Azoury, Omer Kneller, Shaked Rozen, Barry~D Bruner, Alex Clergerie, Yann
  Mairesse, Baptiste Fabre, Bernard Pons, Nirit Dudovich, and Michael
  Kr{\"u}ger.
\newblock Electronic wavefunctions probed by all-optical attosecond
  interferometry.
\newblock {\em Nature Photonics}, 13(1):54--59, 2019.

\bibitem{Dane2012}
D.~E. Laban, A.~J. Palmer, W.~C. Wallace, N.~S. Gaffney, R.~P. M. J.~W.
  Notermans, T.~T.~J. Clevis, M.~G. Pullen, D.~Jiang, H.~M. Quiney, I.~V.
  Litvinyuk, D.~Kielpinski, and R.~T. Sang.
\newblock Extreme ultraviolet interferometer using high-order harmonic
  generation from successive sources.
\newblock {\em Phys. Rev. Lett.}, 109:263902, 2012.

\bibitem{Wang2016}
X.~Wang, H.~Xu, A.~Atia-Tul-Noor, B.~T. Hu, D.~Kielpinski, R.~T. Sang, and
  I.~V. Litvinyuk.
\newblock Isotope effect in tunneling ionization of neutral hydrogen molecules.
\newblock {\em Phys. Rev. Lett.}, 117:083003, Aug 2016.

\bibitem{He2007}
F.~He, C.~Ruiz, and A.~Becker.
\newblock Absorbing boundaries in numerical solutions of the time-dependent
  schr\"odinger equation on a grid using exterior complex scaling.
\newblock {\em Phys. Rev. A}, 75:053407, May 2007.

\bibitem{Kosloff1986}
Kosloff R. and Tal-Ezer H.
\newblock A direct relaxation method for calculating eigenfunctions and
  eigenvalues of the schrodinger equation on a grid.
\newblock {\em Chemical Physics Letters}, 127:223, June 1986.

\bibitem{Crank1947}
Crank J. and Nicolson P.
\newblock A practical method for numerical evaluation of solutions of partial
  differential equations of the heat-conduction type.
\newblock {\em Math. Proc. Cambridge Philos. Soc.}, 43:50, January 1947.

\bibitem{Lein2002}
M.~Lein, N.~Hay, R.~Velotta, J.~P. Marangos, and P.~L. Knight.
\newblock Role of the intramolecular phase in high-harmonic generation.
\newblock {\em Phys. Rev. Lett.}, 88:183903, Apr 2002.

\end{thebibliography}
\bibliographystyle{unsrt}
\clearpage

\section*{Methods}
\setcounter{equation}{0}
\renewcommand{\theequation}{\arabic{equation}}

\noindent \textbf{Experiment:} The experiment is performed with Quantronix Ti:Sapphire laser system consisting of Ti-light modelocked oscillator and Odin-II chirp pulsed amplifier delivering 3 mJ, 37 fs pulses centred at 800 nm wavelength at 1 kHz repetition rate. The laser pulse is coupled into  Neon gas filled hollow core fiber that spectrally broadens the pulse by self-phase modulation. A pair of dielectric chirp compensation mirrors are then used to introduce negative GDD to generate 9 fs pulses. The laser beam is then focused into the HHG chamber by a 750 mm focal length spherical mirror, which leads to an estimated intensity of $\sim$ 5.0$\times$ 10$^{14}$ W/cm$^2$ in the HHG interaction region.\\
 
The Gouy phase interferometer consists of two spatially separated gas jets; fixed bottom jet pointing up towards the laser focus and moveable top jet pointing down before the focus that can translate along all three axes. Each jet is connected with two electromagnetically actuated pulsed microvalves, so that the gases in the jet can be switched according to the measurement requirement. H$_2$ and D$_2$ gases are supplied from separate cylinders connected to the pulsed microvalves.\\

Synchronization of microvalve opening delay with the  laser pulse reduces the background pressure in the HHG generation and detection chambers, which facilitates better signal to noise ratio (for details, see \cite{Mustary2018}). The switching of gases in the jets allows faster measurement and thus minimizing the effect due to laser fluctuation over time and also compensates for a small difference of intensities (and associated phases) at the two different positions along the laser focus. The inner diameter of the jet is only 200 $\mu m$, which is very thin compared to the Rayleigh length of the driving laser beam. The schematic of full experimental setup is shown in the supplementary Fig \ref{Fig 1.pdf}. More details about the instrumentation can be found in \cite{Mustary2018}.\\

\noindent\textbf{Derivation of phase shift:}
The electric field of the HHG signal generated from H$_2$ gas in top jets in figure \ref{Fig 1.pdf} for the $q^{th}$ harmonic order is

\begin{equation}
\vec E_{H_2}=E_{H_2}e^{i(q\omega t+q\phi_{Gouy1}+\phi_{H_2})},
\label{Eq:Ch4:H2_phase}
\end{equation}

here $E_{H_2}$ is maximum field amplitude, $q$ is the harmonic order, $\omega$ is the angular frequency of the driving laser, $\phi_{Gouy1}$ and $\phi_{H_2}$ are the Gouy phase at the top jet position and intrinsic phase of H$_2$ molecule, respectively. Similarly, the electric field of the HHG signal generated from bottom jet with D$_2$ gas is
\begin{equation}
\vec E_{D_2}=E_{D_2}e^{i(q\omega t+q\phi_{Gouy2}+\phi_{D_2})},
\label{Eq:Ch4:D2_phase}
\end{equation}
where $E_{D_2}$ is maximum field amplitude, $\phi_{Gouy2}$ and $\phi_{D_2}$ are the Gouy phase at bottom jet position and intrinsic phase of D$_2$ molecules, respectively. If both jets are ON, two HHG pulses will be generated from top ($H_2$) and bottom ($D_2$) jet with single laser pulse and the total field will be the coherent sum of these two electric fields of equation \ref{Eq:Ch4:H2_phase} and \ref{Eq:Ch4:D2_phase}
\begin{equation}
\vec E_{H_2D_2} =\vec E_{H_2}+\vec E_{D_2}.
\end{equation}
In terms of intensity, the irradiance of the sum of these electric fields is
\begin{equation}
\begin{split}
I_{H_2D_2} & =I_{H_2}+I_{D_2}+2\sqrt{I_{H_2}I_{D_2}}cos(q\Delta\phi_{Gouy}+\Delta\phi_{H_2-D_2}),
\end{split}
\end{equation}
where $\Delta\phi_{Gouy}=\phi_{Gouy1}-\phi_{Gouy2}$ and $\Delta\phi_{H_2-D_2}=\phi_{H_2}-\phi_{D_2}$. Data is then normalized to the in-phase sum of the HHG signals from the H$_2$ jet and D$_2$ jet
\begin{equation}
N_{H_2D_2}=\dfrac{I_{H_2}+I_{D_2}+2\sqrt{I_{H_2}I_{D_2}} cos(q\Delta\phi_{Gouy}+\Delta\phi_{H_2-D_2})}{I_{H_2}+I_{D_2}+2\sqrt{I_{H_2}I_{D_2}}}.
\label{Eq:Ch4:IH2D2}
\end{equation}

\setcounter{figure}{0}
\begin{figure*}[ht!]
   \centering
   \includegraphics[scale=0.45]{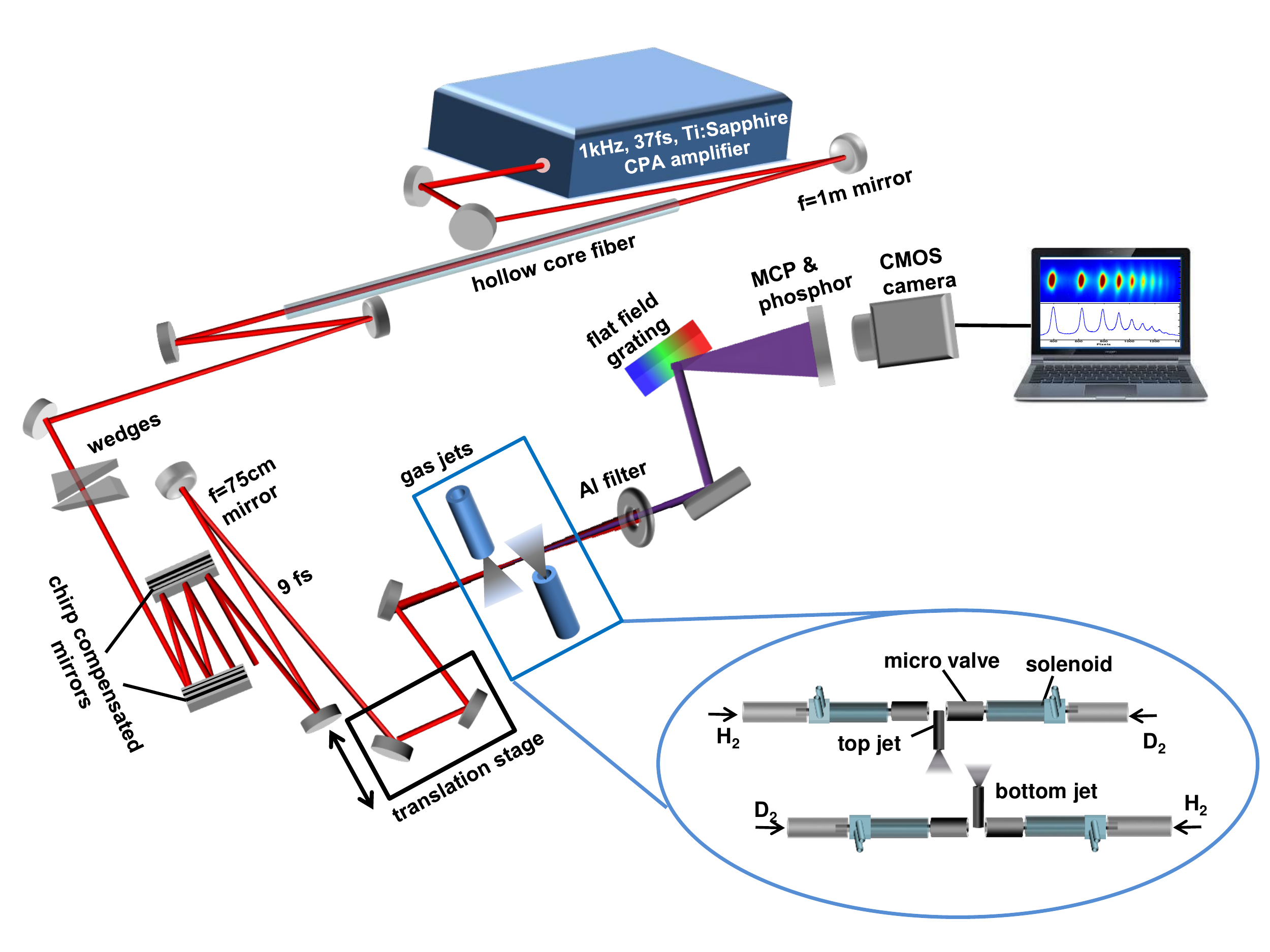}
   \caption{\small {Schematic of the experimental set-up. The ultrashort laser system produces linearly polarized 9 fs pulses, centred at 800 nm and focused down to the HHG chamber by a 750 mm spherical mirror. Two coherent high harmonic pulses are generated from two spatially separated (along laser propagation) gas jets with H$_2$ in one jet and D$_2$ in another. The fundamental IR beam gets blocked by aluminum filter and only HHG beam propagates and spectrally dispersed by the flat field grating spectrometer. The signal is then intensified by MCP detector and finally captured with a CMOS camera. The micro valves (highlighted) are operated in the pulsed mode and synchronized with the laser pulses. The gases are supplied from individual cylinder to the microvalves and then jets. The sequencing, opening and closing time of the valves are synchronized with 1 kHz repetition rate laser pulses by LabVIEW FPGA and microvalve modular controller.}}
   \label{fig:Schematic_Experiment}
   \end{figure*}
When the ordering of gases between top and bottom jets is swapped, i.e. D$_2$ is in the top gas jet and H$_2$ in the bottom gas jet, then the normalized intensity will be

\begin{equation}
N_{D_2H_2}=\dfrac{I_{H_2}+I_{D_2}+2\sqrt{I_{H_2}I_{D_2}} cos(q\Delta\phi_{Gouy}+\Delta\phi_{D_2-H_2})}{I_{H_2}+I_{D_2}+2\sqrt{I_{H_2}I_{D_2}}}.
\label{Eq:Ch4:ID2H2}
\end{equation}
The sign of the molecular phases, $\Delta\phi_{H_2-D_2}$ and $\Delta\phi_{D_2-H_2}$ will be opposite in this case,i.e. $\Delta\phi_{H_2-D_2}=-\Delta\phi_{D_2-H_2}$. Finally, by subtracting  eq(5) from eq(6), the relative high harmonic phase difference between H$_2$ and D$_2$ is expressed as

\begin{equation}
\Delta \phi_{H_2-D_2}=sin^{-1}\Bigg[\dfrac {I_{H_2}+I_{D_2}+2\sqrt{I_{H_2}I_{D_2}}}{4\sqrt{I_{H_2}I_{D_2}}}\Big(\dfrac{\Delta I_N}{sin(q\Delta \phi_{Gouy})}\Big)\Bigg],
\end{equation}
where $I_{H_2}$ and $I_{D_2}$ are HHG yield with H$_2$ and D$_2$ gas respectively. $\Delta I_N=N_{D_2H_2}-N_{H_2D_2}$ and $\Delta \phi_{Gouy}$ is Gouy phase difference of the driving laser.\\

\noindent \textbf{Data analysis:} The acquired data is processed and analyzed by using MATLAB. For each spectra shown in Fig. \ref{fig:Phase_shift}(a) in the main text, the data points around each harmonic peak are plotted. The peak intensity is determined by fitting these data points by a Gaussian function using the least square fitting and the fitting function is
\begin{equation}
f(x)=a \exp^{\Bigg(\dfrac{-(x-b)^2}{2c^2}\Bigg)}
\end{equation}
where $a, b$ and $c$ are the fitting parameters. The peak amplitude of the harmonic signal is given by the fitting parameter $a$. The normalized intensities, as described by equations \ref{Eq:Ch4:IH2D2} and \ref{Eq:Ch4:ID2H2}, are calculated from these peaks.\\

The fitting procedure estimates $1\sigma $ $(68\%)$ confidence bound to quantify the uncertainties, which are used to calculate the standard error. Since $N_{H_2D_2}$ is a function of $I_{H_2}, I_{D_2}, I_{H_2D_2}$ and corresponding error is the combined uncertainty of these $\delta N_{H_2D_2}(\delta I_{H_2},\delta I_{D_2},\delta I_{H_2D_2})$, therefore the error bar of $\Delta N_{{D_2-H_2}}$ is obtained by the propagation of errors and is defined as $\delta (\Delta N_{{D_2}-{H_2}})=\sqrt{(\delta N_{H_2D_2}^2 +\delta N_{D_2H_2}^2)}$.\\

The effective $\Delta \phi_{Gouy}$ for each harmonic order is obtained from the normalized intensity vs separation data of Fig. \ref{Fig 2.pdf}. These data are fitted with a decaying cosine function as
\begin{equation}
f(x)=W\exp^{-Yx}cos(Zx)+C
\end{equation}
where $W$, $Y$, $Z$ and $C$ are the fitting parameters and the $Zx$ gives the $q \Delta \phi_{Gouy}$. Finally, the total error bar of phase shift between H$_2$ and D$_2$ is the propagation error, defined as $\delta\phi_{{D_2}-{H_2}}=\sqrt{\delta_{\Delta N_{{D_2}-{H_2}}}^2+\delta_{sin(q\Delta \phi_{Gouy})}^2}$.\\

%
\noindent\textbf{TDSE simulations:}
We numerically simulate the four dimensional time-dependent Schr\"odinger equation (4D-TDSE) with the single-active-electron (SAE) approximation for H$_2$ and D$_2$ (Hartree atomic units are used $e = m = \hbar = 1$ unless stated otherwise),
\begin{equation}
i\frac {\partial}{\partial t}\Psi(\textbf{\textit{r}},\textbf{\textit{R}};t)=\left[T+V(\textbf{\textit{r}},\textbf{\textit{R}})   \right]\Psi(\textbf{\textit{r}},\textbf{\textit{R}};t),
\end{equation}
where \textbf{\textit{r}} = (\textit{x}, \textit{y}) and \textbf{\textit{R}} = ($\rho$, $\theta$) are  the electronic coordinate and the internuclear displacement, respectively.
The total kinetic energy operator is given as in the velocity gauge
\begin{equation}
T=\frac{[\textbf{\textit{p}}_\textbf{\textit{r}}+\textbf{\textit{A}}(t)]^2}{2}+\frac{\textbf{\textit{p}}_\textbf{\textit{R}}^2}{2\mu}.
\end{equation}
Here, $\mu$ is the reduced nuclear mass, and the laser vector potential is $\textbf{A}(t)=-\int_0^t \textbf{E}(t')dt'$
with $\textbf{E}(t')$ being the electric field vector. $\textbf{\textit{p}}_\textbf{\textit{r}}$ and $\textbf{\textit{p}}_\textbf{\textit{R}}$ are the electronic canonical momentum operator and the relative nuclear momentum operator, respectively.\\


This model reproduces molecular potential energy curves \cite{Lein2005} and thus it is expected to capture the main features of nuclear dynamics of H$_2$ and D$_2$ in laser fields. The interaction between the nuclei and the laser field vanishes because of the homonuclear diatomic molecule.
The effective potential energy is expressed as
\begin{equation}\label{potential energy}
V(\textbf{\textit{r}}, \textbf{\textit{R}}) =V_{BO}^{+}(R)-\sum_{j=1,2}\frac{Z(R,|\textbf{\textit{r}}-\textbf{\textit{R}}_j|)}{\sqrt{|\textbf{\textit{r}}-\textbf{\textit{R}}_j|^2+0.5}},
\end{equation}
which describes the interaction between the outer active electron and the atomic core. $V_{BO}^{+}(R)$ is the lowest isotropic Born-Oppenheimer approximation potential energy surface 1$s\sigma_g$ of H$_2^+$ and D$_2^+$. \textit{R}~=~$|\textbf{\textit{R}}|$ is the internuclear distance  and $\textbf{\textit{R}}_j=\textbf{\textit{R}}/2$ is the $j$-th nuclear displacement in the center of molecular mass coordinate. $Z(R,|\textbf{\textit{r}}-\textbf{\textit{R}}_j|)=\{1+\textrm{exp}[-(\textbf{\textit{r}}-\textbf{\textit{R}}_j)^2/\sigma^2(R)]\}$ is introduced as an effective nuclear charge, which includes the average screening of the nuclei by the second electron with the screening parameter $\sigma(R)$. By choosing proper $\sigma(R)$, the resulting potential energy surface based on Eq. \ref{potential energy} can match the real electronic ground energy of H$_2$ and D$_2$. The other details about this potential can been found in \cite{Lein2005}.\\

In our simualtions, to suppress the unphysical reflection from boundaries and eliminate the long trajectory contributions to high harmonics, a cos$^{1/8}$ masking function has been adopted \cite{He2007}. The initial vibration-rotation ground state  is obtained using the imaginary time propagation \cite{Kosloff1986}, and the Crank-Nicolson method is used to propagate the molecular wave packet in real time \cite{Crank1947}. The harmonic spectrum is calculated via the Fourier transformed dipole acceleration
\begin{equation}
F_x(\omega)=|\frac{1}{(2\pi)^{3/2}}\int D_x(t) e^{-i\omega t} dt|^2
\label{fft}
\end{equation}
with the dipole acceleration
\begin{equation}
D_x(t)=\langle\Psi(\textbf{\textit{r}},\textbf{\textit{R}};t)\left|-\frac{\partial V(\textbf{\textit{r}},\textbf{\textit{R}})}{\partial x}\right |\Psi(\textbf{\textit{r}},\textbf{\textit{R}};t)\rangle.
\label{dipole}
\end{equation}
Here we extract the phase of the complex integral in Eq. \ref{fft} as the phase of the emitted harmonic radiation \cite{Lein2002}. \\

In addition to the 4D calculations, we also performed the reduced 3D calculations by omitting the dynamical terms related to $\theta$ in the 4D model. This can be done because the molecular rotation can be neglected during the harmonic generation. Such a 3D model may directly reveal the contribution from the molecule aligned at a certain $\theta_r$ based on
the $\theta_r$-parameterized molecular wave packet. In all 3D calculations, all parameters and procedures are the same as those implemented in 4D calculations, which can be found in the supplementary information.\\

\section*{Supplementary information}

\noindent\textbf{The calculation parameters:}

\noindent Here, the used laser field polarized along the $\textit{x}$-axis is written as

\setcounter{equation}{0}
\begin{equation}
E(t)=E_0 \textrm{sin}^2(\pi t/\tau)\textrm{cos}(\omega t+\phi).
\end{equation}
$E_0$, $\tau$, $\omega$, and $\phi$ denote the amplitude, the total duration, the angular frequency, and the carrier envelope phase (CEP) of the laser pulse, respectively. In calculations, the laser pulse has the wavelength 800 nm ($\omega=0.057$ a.u.) and lasts ten optical cycles ($\tau = 10T$) with zero CEP ($\phi=0$), and its peak intensity is $5\times10^{14}$ W/cm$^2$ ($E_0 = 0.12$ a.u.). The spatial steps in simulations are $\Delta x=\Delta y=0.3$ a.u., $\Delta \rho=0.04$ a.u., and $\Delta \theta= 2\pi/300$ and the
time step is $\Delta t=0.05$ a.u.. The 4D simulation box has the corresponding grids points: 250 $\times$ 100 $\times$ 100 $\times$ 300. In 3D calculations, the space grids points are 250 $\times$ 100 $\times$ 100. We obtain the HHG spectra for H$_2$ and D$_2$, as shown in Fig. 1. Benefited from the better vibrational autocorrelation character, D$_2$ can generate more intense harmonics than H$_2$ and their intensity difference increases with the harmonic order \cite{Lein2005}. The convergence of the calculations is tested by using finer spatial and time grids whereby almost identical results are obtained.\\

\setcounter{figure}{0}
\begin{suppfigure}[ht!]
   \centering
   \includegraphics[scale=0.25]{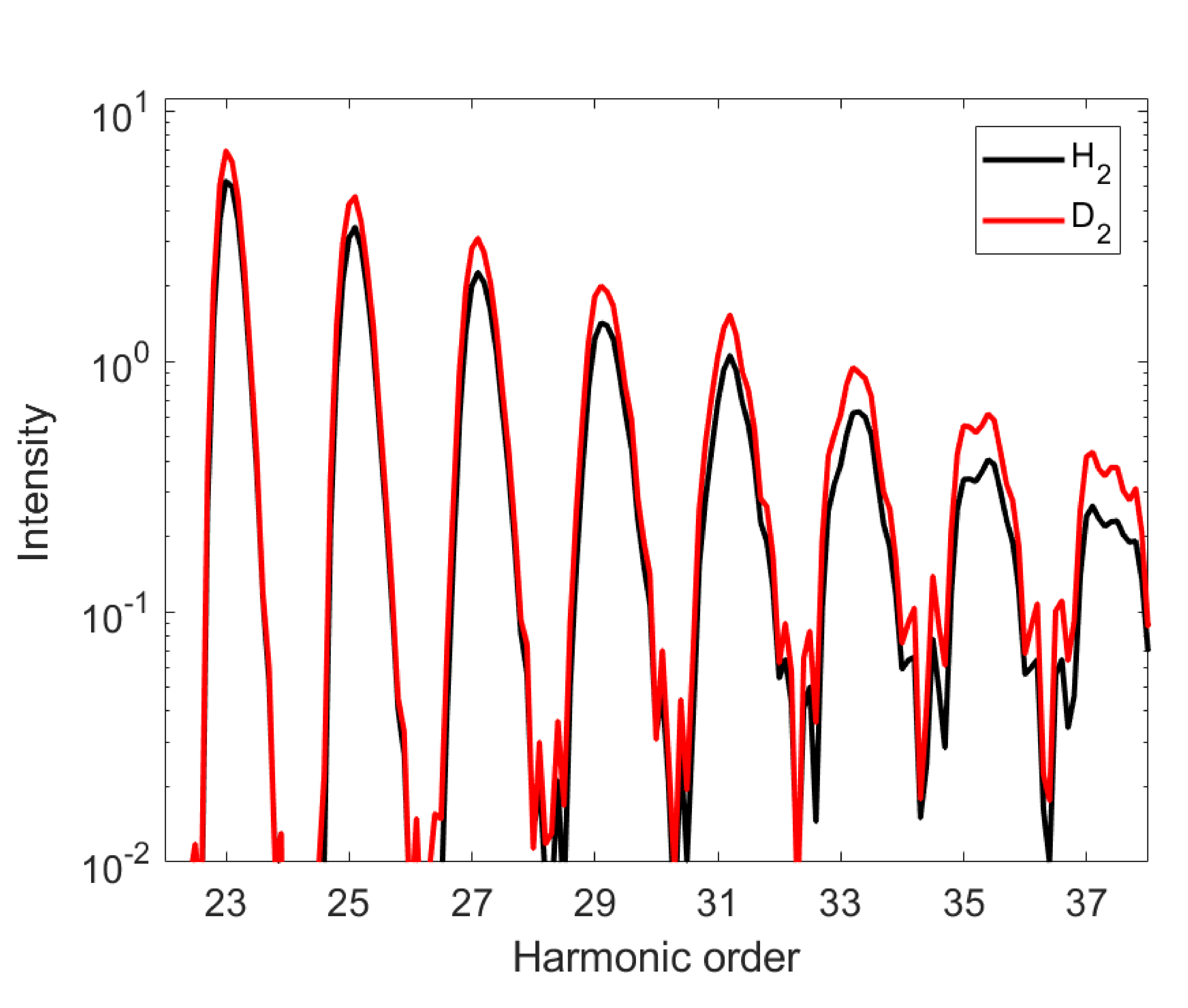}
   \caption{\small {The calculated HHG spectra for H$_2$ (black line) and D$_2$ (red line) from 4D-TDSE simulation.The laser parameters are the wavelength $\lambda$~=~800~nm, the intensity I~=~$5\times10^{14}$ W/cm$^2$, the total duration $\tau$~=~10T, and the sin$^2$ envelope.}}
   \label{Supplementary Fig1}
   \end{suppfigure}

\noindent\textbf{Further analyses based on the Lewenstein model :}
Though the TDSE simulation may quantitatively reproduce the experimental measurements, the Lewenstein model is able to analysis the problem transparently. The time-dependent dipole moment is expressed as

\begin{equation}
\begin{split}
D(t)=2i\int_0^t dt^\prime E(t^\prime)C(t-t^\prime)
\int\frac{d^3p}{(2\pi)^3}\bar{d}^*[p+A(t)]\\
e^{-i\int_{t^\prime}^t\{[p+A(t^{\prime\prime})]^2/2+I_p\}dt^{\prime\prime}}\bar{d}[p+A(t^\prime)]
+c.c.,
\label{Lewenstein}
\end{split}
\end{equation}
where $p$ and $I_p$ are the electronic canonical momentum and molecular ionization energy, respectively.
$t$ and $t^\prime$ denote the electronic ionization and recombination moments. As described by Eq. \ref{Lewenstein}, the nuclear contribution to HHG is related to the vibrational autocorrelation function
$C(t)=\int \chi(R,t=0)\chi(R,t)dR$ and the other terms represent the electronic contribution. $\chi(R,t)$ is approximately governed by the following 1D-TDSE
\begin{equation}
i\frac{\partial}{\partial t}\chi(R,t)=\left[\frac{p_R^2}{2\mu}+V(R)\right ]\chi(R,t),
\label{auto}
\end{equation}

\begin{suppfigure*}[ht]
   \centering
   \includegraphics[scale=0.65]{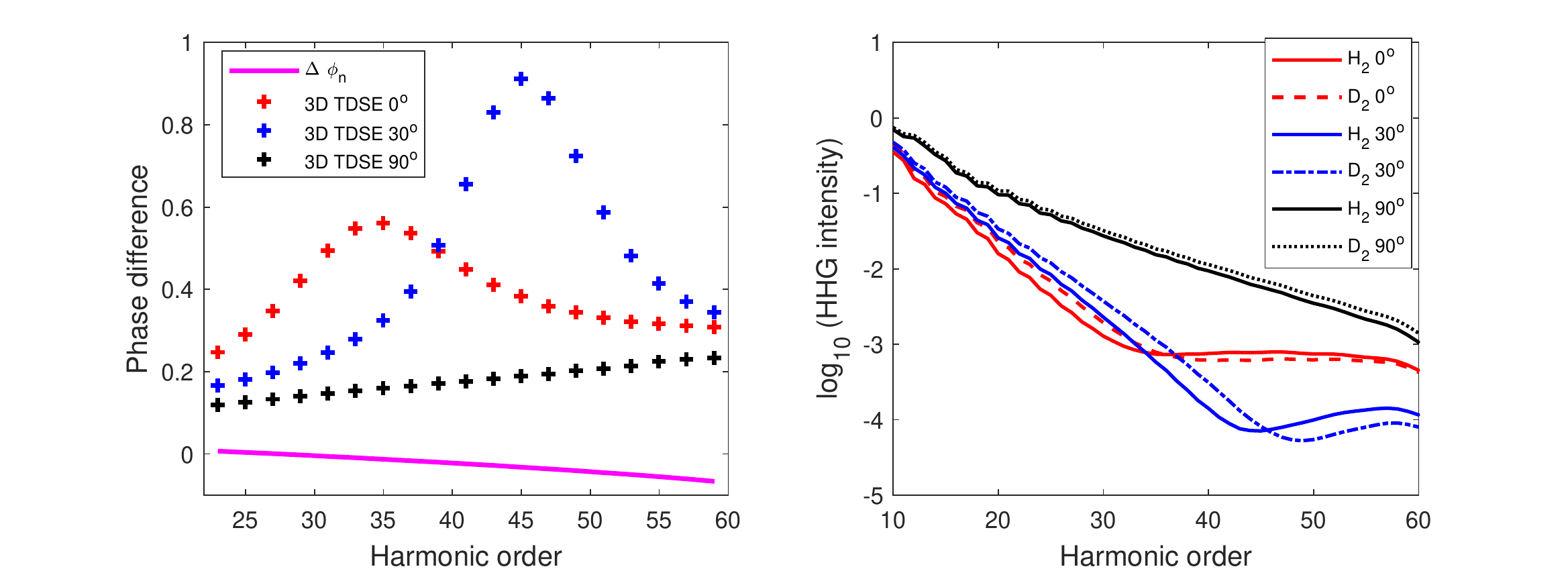}
   \caption{\small {\textbf{(a)} The phase differences between H$_2$ and D$_2$ calculated by 3D-TDSE for the $\theta_r$-aligned molecules with respect to the laser polarization. $\Delta \phi_n$ is the phase difference induced by the nuclear relaxation only. \textbf{(b)} The profiles of high harmonics peaks for H$_2$ and D$_2$ simulated by 3D-TDSE.}}
   \label{Supplementary Fig2}
   \end{suppfigure*}

where $V(R)$ is the potential curve of $1s\sigma_g$ state for H$_2^+$ or D$_2^+$ and \textit{R} is the internuclear distance.\\

Based on Eq. \ref{Lewenstein}, the phase difference must be contributed by the electron dynamics and the nuclear relaxation after the molecular single ionization, as well as the electron-nuclei coupling. We first discuss the the simplest case, that is, $\theta_r=90^{\rm o}$, where the nuclear relaxation and the electronic motion are mutually perpendicular. The phase difference $\Delta \phi_{H_2-D_2}$ is mainly contributed by the nuclear part $\Delta \phi_{n}$ as well as electronic ionization and recombination since the electron dynamical parts in the continuum are almost the same for H$_2$ and D$_2$. To pick up the phase differences induced by the nuclear relaxation, we solely propagate the nuclear wave packets of H$_2^+$ and D$_2^+$ for $\tau_q$ by using Eq. \ref{auto}. Here the time interval $\tau_q$ between ionization and recombination is estimated from the simple-man model. The initial state is the ground nuclear state of H$_2$ or D$_2$ in simulations. The spatial and time steps are $\Delta R = 0.02$ a.u. and $\Delta t = 0.01$ a.u., respectively. The numerical method is the same as those used in 3D or 4D simulations in Methods. The obtained $\Delta \phi_{n}$ shown in Fig. \ref{Supplementary Fig2}(a) is very small about -0.01 radian. The gap between $\Delta \phi_{n}$ and 3D-TDSE results at 90 degrees is up to about 0.2 radian. We inferred that this gap mainly roots in ionization and recombination processes, which is missed in the extended Lewenstein model.\\

Next we turn to other angle cases. Just like Young's double-slit interference in wave optics, when the returning electron momentum $k$ meets the condition $kR$cos$(\theta_r)=\pi$, the high harmonic shows minimal radiation and its phase undergoes an abrupt change at the corresponding frequency $\omega=k^2/2+I_p$. For example, at $\theta_r=0^{\rm o}$ case, the destructive interference mainly changes the phase differences and radiation intensities around the 35th harmonic order, as presented in Fig. \ref{Supplementary Fig2}(a) and (b) respectively, which is estimated as follows. For the 35th order harmonic, the time interval between ionization and rescattering is about 46 a.u. according to the simple-man model under our laser parameters. The nuclear wave packet distributions of H$_2$ and D$_2$ propagated for $\tau_{35}$ by Eq. \ref{auto} have the most conspicuous difference at the nuclear distance $R_c\approx 1.8$ a.u.. The corresponding momentum $k$ and nuclear distance $R_c$ at the rescattering moment right meet the destructive interference condition. Thus, the phase difference at 35th order has the maximal modulation depth with about 0.4 radian compared with $\theta_r=90^{\rm o}$ case. However, the nuclear wave packets of H$_2$ and D$_2$ distribute within a wide $R$ range, so the double-slit interference has also slight effect on the neighbouring order harmonics. Similarly, for $\theta_r=30^{\rm o}$ case, the destructive interference mainly changes the phase differences around 45th harmonic order, as shown in Fig. \ref{Supplementary Fig2}(a), which is beyond our measurement range.

\end{document}